\documentclass[twocolumn,showpacs,preprintnumbers,amsmath,amssymb,APSl,prd,nofootinbib,superscriptaddress]{revtex4-1}

\usepackage{dcolumn}
\usepackage{bm}
\usepackage{ifpdf}
\usepackage{hyperref}
\usepackage{bm}
\usepackage{xcolor,color,graphicx,graphics}
\usepackage[spanish,english]{babel}
\usepackage[latin1]{inputenc}
\usepackage[OT1]{fontenc}
\usepackage{latexsym,amssymb,amsmath,amsfonts}
\usepackage{makeidx}
\usepackage{epsfig,subfigure}
\usepackage{natbib}
\usepackage{epstopdf}
\usepackage{mathrsfs}
\usepackage{hyperref}
\hypersetup{colorlinks=true, linkcolor=blue, citecolor=green}
\usepackage{enumerate}
\usepackage{tikz}

\newcommand*{\rom}[1]{\expandafter\@slowromancap\romannumeral #1@}

\newcommand{\be}{\begin{equation}}
  \newcommand{\ee}{\end{equation}}
\newcommand{\ben}{\begin{eqnarray*}}
  \newcommand{\een}{\end{eqnarray*}}
\newcommand{\bea}{\begin{eqnarray}}
  \newcommand{\eea}{\end{eqnarray}}
\newcommand{\bdm}{\begin{displaymath}}
  \newcommand{\edm}{\end{displaymath}}
\newcommand{\ba}{\begin{align}}
  \newcommand{\ea}{\end{align}}

\makeatother

\definecolor{purple}{rgb}{1,0,1}
\definecolor{lime}{HTML}{A6CE39} 

\newcommand{\orcidicon}{%
	\begin{tikzpicture}
	\draw[lime, fill=lime] (0,0)
		circle [radius=0.16]
		node[white] {{\fontfamily{qag}\selectfont \tiny ID}};
	\draw[white, fill=white] (-0.0625,0.095)
		circle [radius=0.007];
	\end{tikzpicture}	\hspace{-2mm}
}
\newcommand\orcidCabral{{\href{https://orcid.org/0000-0003-2124-5894}{\orcidicon}}}
\newcommand\orcidLobo{{\href{https://orcid.org/0000-0002-9388-8373}{\orcidicon}}}
\newcommand\orcidDiego{{\href{https://orcid.org/0000-0003-3984-9864}{\orcidicon}}}

\begin{document}

\title{Imprints from a Riemann-Cartan space-time on the energy levels of Dirac spinors}

\author{Francisco Cabral\orcidCabral\!\!}
\email{ftcabral@fc.ul.pt}
\affiliation{Instituto de Astrof\'{\i}sica e Ci\^{e}ncias do Espa\c{c}o, Faculdade de
Ci\^encias da Universidade de Lisboa, Edif\'{\i}cio C8, Campo Grande,
P-1749-016 Lisbon, Portugal}
\author{Francisco S. N. Lobo\orcidLobo\!\!} \email{fslobo@fc.ul.pt}
\affiliation{Instituto de Astrof\'{\i}sica e Ci\^{e}ncias do Espa\c{c}o, Faculdade de
Ci\^encias da Universidade de Lisboa, Edif\'{\i}cio C8, Campo Grande,
P-1749-016 Lisbon, Portugal}
\author{Diego Rubiera-Garcia\orcidDiego\!\!} \email{drubiera@ucm.es}
\affiliation{Departamento de F\'isica Te\'orica and IPARCOS, Universidad Complutense de Madrid, E-28040
Madrid, Spain}

\date{\today}

\begin{abstract}
In this work, we investigate the effects of the torsion-fermionic interaction on the energy levels of fermions within a Riemann-Cartan geometry using a model-independent approach. We consider the case of fermions minimally coupled to the background torsion as well as non-minimal extensions via additional couplings with the vector and axial fermionic currents which include parity-breaking interactions. In the limit of zero-curvature, and for the cases of constant and spherically symmetric torsion, we find a Zeeman-like effect on the energy levels of fermions and anti-fermions depending on whether they are aligned/anti-aligned with respect to the axial vector part of the torsion (or to specific combination of torsion quantities), and determine the corresponding fine-structure energy transitions. We also discuss non-minimal couplings between fermionic fields and torsion within the Einstein-Cartan theory and its extension to include the (parity-breaking) Holst term. Finally we elaborate on the detection of torsion effects related to the splitting of energy levels in astrophysics, cosmology and solid state physics using current capabilities.
\end{abstract}

\maketitle

\section{Introduction}

Non-Riemannian geometries naturally appear in gravitation in the gauging (localizing) of space-time symmetries \cite{Blagojevic:2002du,Blagojevic:2012bc,Cabral:2020fax}. A particularly interesting case is the Riemann-Cartan (RC) geometry, linked to the gauging of the Poincar\'{e} group, where in addition to curvature, torsion (the antisymmetric part of the affine connection) is also present. The simplest implementation of the RC geometry is given by the Einstein-Cartan (EC) theory \cite{Hehl:2007bn}, where the corresponding gravitational action is formally given by the Einstein-Hilbert  of General Relativity (GR), but the new freedom encoded in the affine connection allows for new couplings in the matter fields, which are not present when torsion is absent. Analogously to the energy-momentum tensor being the source of curvature, in EC theory the spin tensor feeds the torsion effects into the metric field equations, that become relevant at scales given by Cartan's density. The effects of torsion in bosonic and fermionic fields can be implemented via the covariant derivatives present in the corresponding Lagrangian densities, and manifest at scales different than the Cartan's threshold \cite{Cabral:2019gzh,Cabral:2020mst,Cabral:2020mzw}, while non-minimal couplings can also be considered.

Most research carried out in the corresponding literature so far takes advantage of these new matter couplings to look for new phenomenology in those environments where the spin density is strong enough so as to excite the new dynamics fed by torsion. In this manner, the spin density of fermions translate into modifications of the background space-time metric as compared to the GR predictions. As a consequence, many applications, such as the astrophysics of compact objects or the early and late-time cosmological dynamics have been widely explored and characterized \cite{Boehmer:2008ah,Poplawski:2011jz,Poplawski:2012ab,Vakili:2013fra,Bronnikov:2016xvj,Ivanov:2016xjm,Cembranos:2016xqx,Cembranos:2017pcs,Unger:2018oqo,Mehdizadeh:2018smu}. The main outcome of this analysis is to find new predictions of these theories with torsion that can be compared with their GR counterparts, and that could constitute observational discriminators for the existence of new physics in the strong gravitational field regime. Through this process, these models and their predictions can subsequently be constrained within the newly born field of multimessenger astronomy \cite{Barausse:2020rsu}.

The main aim of this work is to look for more direct physical effects induced by torsion. Indeed, instead of the effects in the space-time background where the torsion fields have been effectively removed in favour of other observables, here we are interested in the behaviour of fermionic fields themselves when propagating on a RC geometry. Let us recall that, in general, the affine connection can be split into its curvature, torsion, and non-metricity pieces (what is known as a metric-affine geometry), allowing for geometrically alternative interpretations but physically equivalent implementations of GR when the lowest scalar object built upon any of these terms is included in the action of the theory (see e.g. \cite{BeltranJimenez:2019tjy} for an interesting discussion on this topic). The interaction of fermionic fields with the space-time geometry in gravitational phenomena is of utmost relevance in this respect, as it can provide a breaking of the degeneracy between these different theories for gravity and space-time \cite{Cabral:2020fax}. Indeed, the different pieces of the affine connection (if non-vanishing) can have direct observable physical effects.

The existence of observable imprints on the modifications to the energy levels on free-falling one-electron atoms induced by curvature in an arbitrary gravitational field are known since the seminal works by Parker \cite{Parker:1980hlc,Parker:1980kw,Parker:1982nk}. On the other hand, the presence of non-metricity has been recently found to yield measurable effects via new 4-fermion contact interactions that can be used to put constraints on the scale of non-metricity, for instance, in electron-positron scattering in particle accelerators \cite{Latorre:2017uve,Delhom:2019wir}. In this paper, we show that both minimal and non-minimal couplings of torsion to Dirac fermions in different backgrounds also yield measurable effects, via a splitting in the energy levels of fermions/antifermions (driven by the axial vector part of the torsion tensor in minimal couplings and also by the trace-vector part for specific non-minimal extensions), depending on the relative orientation between spin and torsion vector/pseudo-vector quantities, which resembles a Zeeman-like effect, allowing for transitions between such different levels\footnote{For further discussions on the experimental manifestations of torsion in the interaction between spinors, torsion, and electromagnetic fields see e.g. \cite{Shapiro1,Shapiro2,Shapiro3,Shapiro4}. }.

This work is organized as follows: In Sec. \ref{sec:II}, we establish the main equations of fermions minimally coupled to the torsion field, and find that they  only interact via the axial vector part of the torsion tensor. In Sec. \ref{sec:III}, we find the energy levels for fermions/anti-fermions in the simplified regime of zero-curvature, taking the ansatze of (static) constant and spherically symmetric torsion, respectively,  and find the frequency of the corresponding transitions. These results are extended in Sec. \ref{sec:IV} to the case of non-minimal couplings between the fermionic currents (vector and axial currents) and the torsion (the axial and trace-vector irreducible parts) where some interaction terms break the parity symmetry. In Sec. \ref{sec:V} we briefly discuss the topic of non-minimal couplings within the EC theory and its generalization to include the parity-breaking Holst term. We conclude in Sec. \ref{sec:VI} with an extended discussion on different types of physical effects connected to the splitting of fermionic energy levels driven by torsion in astrophysics and cosmology.

\section{Torsion-fermions couplings} \label{sec:II}

\subsection{Dirac fermions in a flat space-time}

Let us start our analysis, for self-consistency and self-completeness, by reviewing the corrections to the non-relativistic  Schr\"{o}dinger equation induced by the Dirac equation. Accordingly, consider the Dirac Lagrangian in curved space-time and minimally coupled to the electromagnetic field, given by\footnote{From now on hatted quantities will denote computation with the standard Levi-Civita connection of the curvature (i.e., the Christoffel symbols of the metric).}
\begin{equation}\label{eq:Diraccurve}
\mathcal{\tilde{L}}_{\rm Dirac}=\dfrac{i\hbar}{2}\left(\bar{\psi}\gamma^{\mu}\tilde{D}_{\mu}\psi-(\tilde{D}_{\mu}\bar{\psi})\gamma^{\mu}\psi\right)-m\bar{\psi}\psi+j^{\lambda}A_{\lambda} \ ,
\end{equation}
where $\psi$ denotes spinors and $\bar{\psi}=\psi^+ \gamma^0$ its adjoint, while $j^{\lambda}=q\bar{\psi}\gamma^{\lambda}\psi$ is the $U(1)$ charge current density vector. The (Fock-Ivanenko) covariant derivatives in this space-time are given by
\begin{eqnarray}
\tilde{D}_{\mu}\psi&=&\partial_{\mu}\psi+\dfrac{1}{2}\tilde{w}_{ab\mu}\sigma^{ab}\psi \,, \\
\tilde{D}_{\mu}\bar{\psi}&=&\partial_{\mu}\bar{\psi}-\dfrac{1}{2}\tilde{w}_{ab\mu}\bar{\psi}\sigma^{ab} \,,
\end{eqnarray}
while the matrices $\sigma^{ab}\equiv\tfrac{1}{4}\left[\gamma^{a},\gamma^{b} \right]=\tfrac{1}{2}\gamma^{[a}\gamma^{b]}$ are the Lorentz group generators in the spinorial representation. The effect of space-time curvature is encoded in the Levi-Civita 1-form spin connection, $\tilde{w}_{ab\mu}$, of the Riemann geometry.

The corresponding Dirac equation is obtained by varying the action $S=\tfrac{1}{2\kappa^2} \int d^4x \sqrt{-g} \mathcal{\tilde{L}}_{\rm Dirac}$ of the Dirac Lagrangian (\ref{eq:Diraccurve}) with respect to $\bar{\psi}$ and reads as
\begin{equation}
\label{eq:DiracFockIvaneckoHeisenberg0}
i\hbar \gamma^{\mu}\tilde{D}_{\mu}\psi+(q\gamma^{\mu}A_{\mu}-m)\psi =
0 \ .
\end{equation}
In a flat (Minkowski) space-time, and taking the quasi non-relativistic limit (leaving only terms up to $(v/c)^{2}$), one finds the time independent equation in the static external electromagnetic potential $A_{\mu}=(\phi,\vec{A}\,)$ (here we reinsert the speed of light $c$ for convenience):
\begin{eqnarray}
&&\left[\dfrac{1}{2m}\left(\hat{\vec{p}}-q\vec{A}\right) ^{2}-\dfrac{\hat{p}^{4}}{8m^{3}c^{2}}+q\phi+\dfrac{q\hbar^{2}}{4m^{2}c^{2}}\dfrac{1}{r}\partial_{r}\phi\;\hat{\vec{S}}\cdot\hat{\vec{L}}
 \right. \nonumber \\
&& \left. \hspace{1cm} -\dfrac{q\hbar}{m}\hat{\vec{S}}\cdot\hat{\vec{B}}-\dfrac{q\hbar^{2}}{4m^{2}c^{2}}\partial_{r}\phi\,\partial_{r}-E\right]\psi(\vec{r})=0 \,, \label{eq:Schflat}
\end{eqnarray}
with $E \ll mc^{2}$ and $q\phi \ll mc^{2}$, and spherical symmetry, $\phi=\phi(r)$, is assumed. The solution to these equations gives the four-spinor $\psi=\psi(\vec{r})e^{-iEt/\hbar}$, which corresponds to the eigenfunction of the Hamiltonian with energy $E$. In the expression above, $\hat{\vec{S}}\sim \hbar\vec{\sigma}/2$ is the intrinsic angular momentum (spin), and $\vec{\sigma}=(\sigma^{1},\sigma^{2},\sigma^{3})$ is the Pauli matrices spatial vector. As is well known, the second term on the left-hand side of Eq. (\ref{eq:Schflat}) is a relativistic correction to the three-momentum, the forth and the fifth terms give the spin-orbit and Zeeman-effect magnetic energy, respectively, and the sixth term is the so-called Darwin term correction.

If we consider the case of an electron in the Coulomb potential $\phi=-Ze^{2}/r$, then the corresponding energy levels of this system are given by
\begin{eqnarray}
E&=&mc^{2}\left[1-\dfrac{Z^{2}\alpha^{2}}{2n^{2}}-\dfrac{Z^{4}\alpha^{4}}{2n^{4}}\left(\dfrac{n}{j+1/2}-\dfrac{3}{4}\right)
\right.	\nonumber \\
&& \qquad \qquad\left.
+\mathcal{O}(Z^{6}\alpha^{6})\right] \,,
\end{eqnarray}
where $n$ is the principal quantum number, $j$ is the total angular momentum quantum number, and $\alpha$ is the fine structure constant. The first term correction inside the brackets is the relativistic correction of the energy associated to the mass of the electron, the second term corresponds to the Bohr energy levels, while the next term is the fine structure (spin-orbit) correction. As an example, the fine structure between the energy levels ($nl_{j}$) $2P_{3/2}$ and $2P_{1/2}$ corresponds to an energy difference $\vert\Delta E\vert=mc^{2}Z^{2}\alpha^{4}/32$.

If one considers, instead of the Minkowski limit, the full curved space-time background of a Riemannian geometry, then there will be gravitational metric-induced corrections to the energy levels \cite{Parker:1982nk}, which should become non-negligible for strong gravitational fields. It is thus natural to wonder whether new physical effects will manifest if one generalizes the Riemann geometry to include torsion in a RC space-time. As we will see, in the minimal coupling scenario a clear analogy with the Zeeman-effect term can be recognized.

\subsection{Fermions minimally coupled to torsion}

Let us consider a free Dirac fermionic field minimally coupled to the RC space-time geometry (for a more detailed analysis of fermions in RC and metric-affine geometries see for example \cite{Audretsch:1981xn, Obukhov:2014fta,Obukhov:2018zmz, Adak:2002pq,Cembranos:2018ipn,Bahamonde:2020fnq}). The torsion tensor of the affine connection $\Gamma_{\mu \nu}^{\lambda}$ is defined by $T_{\mu\nu} \equiv \tfrac{1}{2}(\Gamma_{\mu\nu}^{\lambda}-\Gamma_{\nu\mu}^{\lambda})$,
with the irreducible components
$T^{\lambda}_{\;\mu\nu}=\bar{T}^{\lambda}_{\;\mu\nu}+\tfrac{2}{3}\delta^{\lambda}_{[\nu}T_{\mu]}+g^{\lambda\sigma}\epsilon_{\mu\nu\sigma\rho} \breve{T}^{\rho}$, and the traceless tensor obeying $\bar{T}^{\lambda}_{\;\mu\lambda}=0$ and $\epsilon^{\lambda\mu\nu\rho}\bar{T}_{\mu\nu\rho}=0$, while $T_{\mu}$ is the trace vector and
\begin{equation} \label{eq:psaxv}
\breve{T}^{\lambda}\equiv \frac{1}{6}\epsilon^{\lambda\alpha\beta\gamma}T_{\alpha\beta\gamma} \ ,
\end{equation}
is the  pseudo-trace (axial) vector, which will play a key role in our work.

The minimally-coupled fermionic (Dirac) piece, Eq. (\ref{eq:actionEC}), in presence of torsion, is given by
\begin{equation}\label{eq:mattfer}
\mathcal{L}_{\rm Dirac}=\dfrac{i\hbar}{2}\left(\bar{\psi}\gamma^{\mu}D_{\mu}\psi-(D_{\mu}\bar{\psi})\gamma^{\mu}\psi\right)-m\bar{\psi}\psi \ ,
\end{equation}
which is formally equal to Eq. (\ref{eq:Diraccurve}), but where the generalized Fock-Ivanenko covariant derivatives of spinors are now defined as
\begin{equation}
D_{\mu}\psi=\partial_{\mu}\psi+\dfrac{1}{2}w_{ab\mu}\sigma^{ab}\psi=\tilde{D}_{\mu}\psi+\dfrac{1}{4}K_{ab\mu}\gamma^{[a}\gamma^{b]}\psi \label{eq:cov1}  \ ,
\end{equation}
and
\begin{equation}
D_{\mu}\bar{\psi}=\partial_{\mu}\bar{\psi}-\dfrac{1}{2}w_{ab\mu}\bar{\psi}\sigma^{ab}=\tilde{D}_{\mu}\bar{\psi}-\dfrac{1}{4}K_{ab\mu}\bar{\psi}\gamma^{[a}\gamma^{b]} \label{eq:cov2} \ ,
\end{equation}
where the Lorentzian spin connection ($w_{ab\nu}=-w_{ba\nu}$) of the RC space-time can be written as the spin connection of the Riemann geometry plus the so-called contortion tensor $K_{\alpha\mu\nu} \equiv T_{\alpha\mu\nu}+2T_{(\mu\nu)\alpha}$, that is
\begin{equation}
w_{ab\mu}=\tilde{w}_{ab\mu}+K_{ab\mu} \ .
\end{equation}
Replacing the expressions of the covariant derivatives (\ref{eq:cov1}) and (\ref{eq:cov2}) in the Lagrangian density (\ref{eq:mattfer}) yields
\begin{equation}
\mathcal{L}_{\rm Dirac}=\tilde{\mathcal{L}}_{\rm Dirac}+\dfrac{i\hbar}{8}K_{ab\mu}\bar{\psi}\lbrace\gamma^{\mu},\gamma^{a}\gamma^{b}\rbrace\psi \ ,
\end{equation}
where $\tilde{\mathcal{L}}_{\rm Dirac}$ is given by Eq. (\ref{eq:Diraccurve}).

Using the canonical properties of the tetrads we can write the contractions $K_{ab\mu}=\vartheta^{c}_{\,\,\mu}K_{abc}$ and $\gamma^{\mu}=e^{\,\,\mu}_{d}\gamma^{d}$, so that the Lagrangian density above can also be written as
\begin{equation}
\mathcal{L}_{\rm Dirac}=\tilde{\mathcal{L}}_{\rm Dirac}+\dfrac{i\hbar}{8}K_{abc}\bar{\psi}\lbrace\gamma^{c},\gamma^{a}\gamma^{b}\rbrace\psi \ .
\end{equation}
Next, by using the identities $\lbrace\gamma^{c},\gamma^{a}\gamma^{b}\rbrace=2\gamma^{[c}\gamma^{a}\gamma^{b]}=-2i\epsilon^{cabd}\gamma_{d}\gamma^{5}$ we can rewrite it as
\begin{equation}
\mathcal{L}_{\rm Dirac}=\tilde{\mathcal{L}}_{\rm Dirac}+
\dfrac{i\hbar}{4}K_{abc}\bar{\psi}\gamma^{[c}\gamma^{a}\gamma^{b]}\psi \ ,
\end{equation}
and noting that $K_{[\alpha\beta\gamma]}=T_{\alpha\beta\gamma}$, we arrive at the final expression
\begin{equation} \label{eq:diraclagfull}
\mathcal{L}_{\rm Dirac}=\tilde{\mathcal{L}}_{\rm Dirac}+3\breve{T}^{\lambda}\breve{s}_{\lambda} \ ,
\end{equation}
where
\begin{equation} \label{eq:Daspc}
\breve{s}^{\lambda}\equiv \dfrac{\hbar}{2} \bar{\psi}\gamma^{\lambda}\gamma^{5}\psi \ ,
\end{equation}
is the Dirac axial spin vector current. Note that in this expression we have reinserted the space-time (holonomic) indices. This simple expression, which is valid for any Dirac field minimally coupled to a RC space-time geometry (regardless of the gravitational theory chosen) means that, in the minimal coupling case, Dirac fermionic fields only interact with the axial vector part of torsion.

The  axial vector $\breve{s}^{\lambda}$ in Eq. (\ref{eq:Daspc})  can be understood as representing the density of fermionic spin (spin/volume or energy/area, in $c=1$ units). To see this more explicitly, let us consider the $\gamma^{a}\gamma^{5}$ matrices for $a=0,1,2,3$, that is
\begin{equation}
\gamma^{a}\gamma^{5}= \Bigg\{\begin{pmatrix}
 0 & I\\
-I & 0
\end{pmatrix},
\begin{pmatrix}
 \sigma^{i} & 0\\
0 & -\sigma^{i}
\end{pmatrix}
\Bigg \} \ ,
\end{equation}
(with $i=1,2,3$), respectively, and with $I$ representing the $2 \times 2$ identity matrix.

Since the eigenvalues of the Pauli matrices are $\lambda=\pm 1$ for the spin up/down configurations, and using the fact that in the usual Pauli-Dirac representation the $\sigma^{3}$ matrix is already diagonal, we can use this direction as the one relative to which we define the up and down spin states. Then one can show that
\begin{equation}
\vert\breve{s}^{3}\vert\sim\dfrac{\hbar}{2}n \ ,
\end{equation}
where $n$ is a normalization constant giving the number of particles (or anti-particles) per volume.

The torsion-spin interaction term in Eq. (\ref{eq:diraclagfull}) actually resembles a Zeeman-like effect with the axial spin vector playing the role of an external magnetic field:
\begin{equation}
\mathcal{L}_{\rm ts} \sim \breve{\bold{T}} \cdot \breve{\bold s} \,.
\end{equation}
where bold letters indicate a product of two (axial) vectors.

We can now find the Dirac equation corresponding to the Lagrangian density (\ref{eq:diraclagfull}) for spinors and adjoint spinors as
\begin{eqnarray}
i\hbar \gamma^{\mu}\tilde{D}_{\mu}\psi-m\psi &=&
-\dfrac{3\hbar}{2} \breve{T}^{\lambda}\gamma_{\lambda}\gamma^{5}\psi \,, \label{eq:DiracFockIvaneckoHeisenberg}\\
i\hbar (\tilde{D}_{\mu}\bar{\psi})\gamma^{\mu}+m\bar{\psi} &=&
-\dfrac{3\hbar}{2} \breve{T}^{\lambda}\bar{\psi}\gamma_{\lambda}\gamma^{5} \,, \label{eq:DiracFockIvaneckoHeisenbergadjoint}
\end{eqnarray}
respectively. In the next section we shall study specific solutions of this system in order to determine the energy levels for fermions and anti-fermions.

\section{Imprints of torsion upon fermion/anti-fermion energy levels}  \label{sec:III}

\subsection{Constant background axial torsion}

We start our analysis by taking again the zero-curvature limit and, moreover, we specify an axial torsion vector along one specific direction (for example the $z$ axis of a cartesian coordinate system). Under these conditions, the Dirac equation (\ref{eq:DiracFockIvaneckoHeisenberg}) reads
\begin{equation}
\label{eq:DiracFockIvaneckoHeisenbergflatlimit}
i\hbar \gamma^{\alpha}\partial_{\alpha}\psi=m\psi
-\dfrac{3\hbar}{2} \breve{T}^{3}\gamma_{3}\gamma^{5}\psi \,.
\end{equation}
More explicitly, using the expressions of the Pauli matrices and recalling that $\gamma^{3}=-\gamma_{3}$, one can recast this equation into the dynamical system
 \begin{eqnarray}
 i\hbar\Big(\partial_{t}\psi^{I}+\sigma^{k}\partial_{k}\psi^{II}\Big)&=&\left(m+\dfrac{3\hbar}{2}\breve{T}^{3}\sigma^{3}\right)\psi^{I} ,
 	\\
-i\hbar\Big(\partial_{t}\psi^{II}+\sigma^{k}\partial_{k}\psi^{I}\Big)&=&\left(m-\dfrac{3\hbar}{2}\breve{T}^{3}\sigma^{3}\right)\psi^{II} ,
\end{eqnarray}
%
where we have introduced the notation $\psi^I=(\psi^1,\psi^2)$ and $\psi^{II}=(\psi^3,\psi^4)$, while $k=1,2,3$, and the $2\times 2$ identity matrix $I$ is implicit in the first terms of the left-hand side and in the first (mass) terms on the right-hand side.

From these equations one can see that the axial-axial interaction between the fermionic spin density and the background space-time torsion gives a spin-dependent energy (depending on the relative orientation between the axial spin vector and the background space-time torsion). Therefore, an electron or any massive free fermion in a well defined momentum (eigen)state will have two possible energy levels depending on the alignment/anti-alignment between its spin and the axial torsion vector, which is analogous to the Zeeman effect. Moreover, if we assume that $\breve{T}^{3}>0$, then the anti-alignment is preferred for the fermion as it corresponds to the lower energy level and the same result occurs for the anti-fermion.

To make our analysis more concrete, let us assume the simpler case of a static, constant torsion field. Consider then a 4-spinor $\psi=\psi(\vec{r}\,)e^{-iEt/\hbar}$, corresponding to the eigenfunction of a well-defined energy state. After substituting in Eq. (\ref{eq:DiracFockIvaneckoHeisenbergflatlimit}), we obtain the time-independent equation
\begin{equation}
\label{eq:Diracflatlimittimeind}
-i\hbar \gamma^{k}\partial_{k}\psi+\Big(m
-\dfrac{3\hbar}{2} \breve{T}^{3}\gamma_{3}\gamma^{5}\Big)\psi(\vec{r})=\gamma^{0}E\psi(\vec{r}) \ .
\end{equation}
In terms of their components this equation reads
\begin{eqnarray}
-i\hbar\sigma^{k}\partial_{k}\psi^{II}&=&\left(E-m-\dfrac{3\hbar}{2}\breve{T}^{3}\sigma^{3}\right)\psi^{I}, \\
-i\hbar\sigma^{k}\partial_{k}\psi^{I}&=&\left(E+m-\dfrac{3\hbar}{2}\breve{T}^{3}\sigma^{3}\right)\psi^{II} \,.
\end{eqnarray}

Moreover, taking into account the harmonic solution $\psi(\vec{r})=\chi e^{i\vec{k}\cdot\vec{r}}=\chi e^{i\vec{p}\cdot\vec{r}/\hbar}$, corresponding to a well defined momentum state, where $\chi$ is a constant 4-spinor, we get the system of equations for the $\chi$ components as
\begin{eqnarray}
p_{1}\chi^{4}-ip_{2}\chi^{4}+p_{3}\chi^{3}&=&\left(E-m-\dfrac{3\hbar}{2}\breve{T}^{3}\right)\chi^{1} , \nonumber \\
p_{1}\chi^{3}+ip_{2}\chi^{3}-p_{3}\chi^{4}&=&\left(E-m+\dfrac{3\hbar}{2}\breve{T}^{3}\right)\chi^{2} , \nonumber \\
-\left(-p_{1}\chi^{2}+ip_{2}\chi^{2}-p_{3}\chi^{1}\right)&=&\left(E+m-\dfrac{3\hbar}{2}\breve{T}^{3}\right)\chi^{3} , \nonumber \\
-\left(-p_{1}\chi^{1}-ip_{2}\chi^{1}+p_{3}\chi^{2} \right)&=&\left(E+m+\dfrac{3\hbar}{2}\breve{T}^{3}\right)\chi^{4}  , \nonumber
\end{eqnarray}
respectively.

Since $\chi$ is assumed to have constant components, the background torsion itself has to be constant too.
In this static, constant background axial torsion regime, assuming again that torsion is positively oriented, $\breve{T}^{3}>0$, there are two independent solutions for the spinor $\psi(\vec{r},t)=\chi e^{i\left(\vec{p}\cdot\vec{r}-Et\right)/\hbar}$, corresponding to the free particle momentum eigenstates with spin up and spin down. But, as opposed to Dirac theory in Minkowski space-time, in this case the presence of torsion breaks the degeneracy in energy and these two states have different (positive) energy values. As an example, consider the case of motion along the $p_{1}$ direction for this eigenstate, for which we get
\begin{eqnarray}
p_{1}\chi^{4}&=&\left(E-m-\dfrac{3\hbar}{2}\breve{T}^{3}\right)\chi^{1} , \nonumber \\
p_{1}\chi^{3}&=&\left(E-m+\dfrac{3\hbar}{2}\breve{T}^{3}\right)\chi^{2} , \nonumber \\
p_{1}\chi^{2}&=&\left(E+m-\dfrac{3\hbar}{2}\breve{T}^{3}\right)\chi^{3} , \nonumber \\
p_{1}\chi^{1}&=&\left(E+m+\dfrac{3\hbar}{2}\breve{T}^{3}\right)\chi^{4} . \nonumber
\end{eqnarray}

The two possible energy solutions for the particle are then given by
\begin{equation}
E_{\pm}^{2}=p^{2}+\left(m\pm\dfrac{3\hbar}{2}\breve{T}^{3}\right)^{2},
\end{equation}
for the spin up/down, respectively. The independent solutions for the spin up (aligned) state and the spin down (anti-aligned) state are given by
\begin{equation}
N\begin{pmatrix}
 1 \\
0 \\
0 \\
\dfrac{p}{E+\left(m + \dfrac{3\hbar}{2}\breve{T}^{3}\right)}
\end{pmatrix} , \quad
N\begin{pmatrix}
0 \\
1 \\
\dfrac{p}{E+\left(m - \dfrac{3\hbar}{2}\breve{T}^{3}\right)} \\
0
\end{pmatrix}, \nonumber
\end{equation}
respectively, where $N$ is a normalization constant (typically chosen to satisfy  $\psi^{\dagger}\psi=2E$) given in this case by $N=\sqrt{E+(m \pm 3\hbar \, \breve{T}^{3}/2)}$ for the spin up/down (aligned/anti-aligned) state. From this discussion, we see that not only the axial-axial torsion-spin interaction is analogous to a Zeeman effect but also the equations reveal that one could think of the fermion state with the spin aligned with torsion as being slightly more massive than the fermion state with the spin anti-aligned to the axial torsion. In the coupling to the space-time structure, torsion is therefore providing an effective mass to fermions that distinguishes between spin states.

Let us also note that in this regime of static constant background torsion there are two more independent solutions for the spinor $\psi(\vec{r})=\chi e^{-i\left(\vec{p}\cdot\vec{r}-Et\right)/\hbar}$, corresponding to the free anti-particle momentum eigenstates with spin down or spin up, respectively. In this case, we obtain
\begin{equation}
N\begin{pmatrix}
 0 \\
\dfrac{p}{E+\left(m - \dfrac{3\hbar}{2}\breve{T}^{3}\right)} \\
1 \\
0
\end{pmatrix}, \quad
N\begin{pmatrix}
\dfrac{p}{E+\left(m + \dfrac{3\hbar}{2}\breve{T}^{3}\right)} \\
 0 \\
0 \\
1
\end{pmatrix},
\end{equation}
respectively, with $N=\sqrt{E+(m \mp 3\hbar\,\breve{T}^{3}/2)}$ for the spin down/up (anti-aligned/aligned) states, respectively. The two corresponding energy levels are
\begin{equation}
E^{2}=p^{2}+\left(m \mp \dfrac{3\hbar}{2}\breve{T}^{3}\right)^{2} \ ,
\end{equation}
for the spin down/up states.

For completeness, let us also mention that in the general case of motion along any direction, with $\vec{p}=(p_{1},p_{2},p_{3})$, then we would reach similar conclusions with the spin up and spin down solutions for particles:
\begin{equation}
N\begin{pmatrix}
 1 \\
0 \\
\dfrac{p_{3}}{E+\left(m+\dfrac{3\hbar}{2}\breve{T}^{3}\right)} \\
\dfrac{p_{1}+ip_{2}}{E+\left(m+\dfrac{3\hbar}{2}\breve{T}^{3}\right)}
\end{pmatrix} , \quad
N\begin{pmatrix}
  0 \\
1 \\
\dfrac{p_{1}-ip_{2}}{E+\left(m-\dfrac{3\hbar}{2}\breve{T}^{3}\right)} \\
\dfrac{-p_{3}}{E+\left(m-\dfrac{3\hbar}{2}\breve{T}^{3}\right)}
\end{pmatrix} ,
\end{equation}
and those for anti-particles:
\begin{equation}
N\begin{pmatrix}
\dfrac{p_{1}-ip_{2}}{E+\left(m+\dfrac{3\hbar}{2}\breve{T}^{3}\right)}\\
\dfrac{-p_{3}}{E+\left(m+\dfrac{3\hbar}{2}\breve{T}^{3}\right)} \\
0 \\
1
\end{pmatrix}, \quad
N\begin{pmatrix}
\dfrac{p_{3}}{E+\left(m-\dfrac{3\hbar}{2}\breve{T}^{3}\right)}\\
\dfrac{p_{1}+ip_{2}}{E+\left(m-\dfrac{3\hbar}{2}\breve{T}^{3}\right)} \\
1 \\
0
\end{pmatrix}  \ .
\end{equation}
In all cases the energy of the anti-aligned state is lower than the aligned state.

Let us denote by $m_{\breve{T}}$ the mass correction due to the spin-torsion interaction, and consider the two possible energy levels $E_{1}$ and $E_{2}$, with $E_{2}>E_{1}$. We therefore get the expression for the energy transition
\begin{equation}
h\nu= E_{2}-E_{1}=\dfrac{4mm_{\breve{T}}}{\tilde{p}_{+}+\tilde{p}_{-}} \ ,
\end{equation}
where $\tilde{p}^2_{\pm}=p^2+(m \pm m_{\breve{T}})^{2}$, and in the reference frame of the particle we obtain
\begin{equation}
h\nu= E_{2}-E_{1}=\dfrac{1}{2}m_{\breve{T}}=\dfrac{3\hbar}{4}\breve{T} \ .
\end{equation}
Therefore, reinserting the speed of light in vacuum, we get
\begin{equation}
\nu=\frac{3c}{8\pi} \breve{T}
\end{equation}
If we consider, for instance, $ \breve{T} \sim 10^{-16}m^{-1}$, then we end up with the prediction of a transition in the $\nu \sim$nHz regime.

\subsection{Spherically symmetric torsion background}

In this section, we analyze the case of a static, spherically symmetric torsion background, which is relevant for astrophysical applications.
The Dirac equations in this case are still given by Eqs. (\ref{eq:DiracFockIvaneckoHeisenberg}) and (\ref{eq:DiracFockIvaneckoHeisenbergadjoint}). To estimate the effect of this scenario on the energy levels we consider the following axial torsion ansatz around some astrophysical source:
\begin{equation}
\breve{T}^{\mu}(r)=b^{\mu}f(r) \ ,
\end{equation}
where $b^{\mu}$ is a constant (axial) 4-vector. If we neglect the effect of curvature, the limit of the generalized Dirac equation above is
\begin{equation}
\label{eq:DiracFockIvaneckoHeisenbergflatlimitsphericaltors}
i\hbar \gamma^{\alpha}\partial_{\alpha}\psi=m\psi
-\dfrac{3\hbar}{2} \breve{T}^{\lambda}(r)\gamma_{\lambda}\gamma^{5}\psi \,.
\end{equation}

The torsion-spin interaction can be seen as a small perturbation to an (unperturbed) time-independent Hamiltonian. Using perturbation theory to first order, we have then
\begin{equation}
E\simeq E_{(0)}+\left<\psi_{(0)}\right| \hat{U}_{ts}\left|\psi_{(0)} \right> \ ,
\end{equation}
where $\psi_{(0)}$ are the eigenstates of the unperturbed Hamiltonian associated to the eigenvalue $E_{(0)}$. Again, taking the 4-spinor $\psi=\psi(\vec{r})e^{-iEt/\hbar}$, corresponding to the eigenfunction of a well-defined energy state, we obtain the time-independent equation
\begin{equation}
\label{eq:Diracflatlimittimeindsphericaltors}
-i\hbar \gamma^{k}\partial_{k}\psi+\Big(m
-\dfrac{3\hbar}{2} \breve{T}^{\lambda}(r)\gamma_{\lambda}\gamma^{5}\Big)\psi=\gamma^{0}E\psi \,,
\end{equation}
such that the torsion-spin operator $\hat{U}_{ts}$ reads
\begin{equation} \label{eq:Utorsion}
\hat{U}_{ts}=-\dfrac{3\hbar}{2} \hat{\breve{T}}^{\lambda}(r)\gamma_{\lambda}\gamma^{5}.
\end{equation}

Now, consider the 4-spinor state
\begin{eqnarray}
\left| \psi_{(0)} \right>&=&\left| \psi_{(0)}^{1} \right>\begin{pmatrix}
 1 \\
0 \\
0 \\
0
\end{pmatrix}+ \left| \psi_{(0)}^{2} \right>\begin{pmatrix}
0 \\
1 \\
0 \\
0
\end{pmatrix}\nonumber  \\
&&+ \left| \psi_{(0)}^{3} \right>\begin{pmatrix}
0 \\
0 \\
1 \\
0
\end{pmatrix}+ \left| \psi_{(0)}^{4} \right>\begin{pmatrix}
0 \\
0 \\
0 \\
1
\end{pmatrix} \ ,
\end{eqnarray}
solution to the unperturbed Hamiltonian. In configuration space, this expression becomes
\begin{eqnarray}
\left<\vec{r} \; \big| \psi_{(0)} \right>&=&\psi_{(0)}^{1}(r)\begin{pmatrix}
 1 \\
0 \\
0 \\
0
\end{pmatrix}+ \psi_{(0)}^{2}(r)\begin{pmatrix}
0 \\
1 \\
0 \\
0
\end{pmatrix} \nonumber \\
&&+\psi_{(0)}^{3}(r)\begin{pmatrix}
0 \\
0 \\
1 \\
0
\end{pmatrix}+\psi_{(0)}^{4}(r)\begin{pmatrix}
0 \\
0 \\
0 \\
1
\end{pmatrix} \,.
\end{eqnarray}
If we assume the motion to take place along a specific direction, then there are four independent solutions, two for the particle states (up/down):
\begin{equation}
N\begin{pmatrix}
 1 \\
0 \\
0 \\
\dfrac{p}{E_{(0)}+m}
\end{pmatrix}e^{i\vec{p}\cdot\vec{r}/\hbar}, \quad
N\begin{pmatrix}
  0 \\
1 \\
\dfrac{p}{E_{(0)}+m} \\
0
\end{pmatrix}e^{i\vec{p}\cdot\vec{r}/\hbar} , \label{eq:supud}
\end{equation}
and two for the anti-particle states (down/up)
\begin{equation}
N\begin{pmatrix}
 0 \\
\dfrac{p}{E_{(0)}+m} \\
1 \\
0
\end{pmatrix}e^{-i\vec{p}\cdot\vec{r}/\hbar} , \quad
N \begin{pmatrix}
\dfrac{p}{E_{(0)}+m} \\
0 \\
0 \\
1
\end{pmatrix}e^{-i\vec{p}\cdot\vec{r}/\hbar} ,
\end{equation}
 with $N=\sqrt{E_{(0)}+m}$ and $E_{(0)}^{2}=p^{2}+m^{2}$.

 Next, we need to compute from Eq. (\ref{eq:Utorsion}) the following quantity
\begin{eqnarray}
\left<\psi_{(0)} \right| \hat{U}_{ts} \left| \psi_{(0)} \right>&=&-\dfrac{3\hbar}{2} \int\psi^{\dagger}_{0}(\vec{r})\hat{\breve{T}}^{\lambda}(r)\gamma_{\lambda}\gamma^{5}\psi_{0}(\vec{r})d^{3}r
	\nonumber \\
\qquad \qquad &=&-\dfrac{3\hbar b_0}{2}  \int f(r)\psi^{\dagger}_{0}(\vec{r})\gamma_{0}\gamma^{5}\psi_{0}(\vec{r})d^{3}r
	\\
\qquad \qquad \quad &+&\dfrac{3\hbar}{2}\sum_{i=1}^{3}b^{i} \int f(r)\psi^{\dagger}_{0}(\vec{r})\gamma^{i}\gamma^{5}\psi_{0}(\vec{r})d^{3}r \nonumber  \ ,
\end{eqnarray}
which, taking into account the quite useful general relations
\begin{eqnarray}
\gamma_{0}\gamma^{5}\begin{pmatrix}
z_{1} \\
z_{2} \\
z_{3} \\
z_{4}
\end{pmatrix}&=&\begin{pmatrix}
z_{3} \\
z_{4} \\
-z_{1} \\
-z_{2}
\end{pmatrix},\qquad \gamma^{1}\gamma^{5}\begin{pmatrix}
z_{1} \\
z_{2} \\
z_{3} \\
z_{4}
\end{pmatrix}=\begin{pmatrix}
z_{2} \\
z_{1} \\
-z_{4} \\
-z_{3}
\end{pmatrix} , \nonumber \\
\gamma^{2}\gamma^{5}\begin{pmatrix}
z_{1} \\
z_{2} \\
z_{3} \\
z_{4}
\end{pmatrix}&=&\begin{pmatrix}
-iz_{2} \\
iz_{1} \\
iz_{4} \\
-iz_{3}
\end{pmatrix},\qquad\gamma^{3}\gamma^{5}\begin{pmatrix}
z_{1} \\
z_{2} \\
z_{3} \\
z_{4}
\end{pmatrix}=\begin{pmatrix}
z_{1} \\
-z_{2} \\
-z_{3} \\
z_{4}
\end{pmatrix} ,\nonumber
\end{eqnarray}
and assuming the particle with the spin up configuration in Eq. (\ref{eq:supud}), with $z_{1}=Ne^{i\vec{p}\cdot\vec{r}/\hbar}$, $z_{4}=N\frac{p}{E+m}e^{i\vec{p}\cdot\vec{r}/\hbar}$ and $z_{2}=z_{3}=0$, we arrive at
\begin{equation}
\left<\psi_{(0)}\right| \hat{U}_{ts}\left|\psi_{(0)}\right>=\dfrac{3\hbar b^{3}}{2}N^{2}\left[1+\dfrac{p^{2}}{(E_{(0)}+m)^{2}}\right] F(r) \ , \label{eq:Usu}
\end{equation}
where $F(r)=b^{3}\int f(r)d^{3}r$ represents the geometrical factor coming from a spherically symmetric torsion function integrated over the relevant volume of the spatial 3-dimensional hypersurfaces, for a specific space-time foliation. Doing the same exercise with the spin down state  with $z_{2}=Ne^{i\vec{p}\cdot\vec{r}/\hbar}$, $z_{3}=N\frac{p}{E_{(0)}+m}e^{i\vec{p}\cdot\vec{r}/\hbar}$ and $z_{1}=z_{4}=0$, we also get Eq. (\ref{eq:Usu}) but with a global minus sign. The energy difference between these two states thus becomes
\begin{equation}
 \delta E   =3\hbar b^{3}N^{2}\left(1+\frac{p^{2}}{(E_{(0)}+m)^{2}}\right) F(r) \ ,
\end{equation}
which corresponds to the frequency
\begin{equation}
\nu= \dfrac{\delta E}{h}=\dfrac{3E_{(0)}}{\pi}F(r) \ ,
\end{equation}
where we have used again the conventional normalization $\psi^{\dagger}_{0}\psi_{0}=2E_{0}$. Note that the same result for the energy levels would have been obtained if we had considered all the components of the 3-momentum.


\section{Non-minimally coupled fermions to torsion}  \label{sec:IV}

Let us study now a fermionic Dirac Lagrangian non-minimally coupled to the RC  geometry. Consider the vector and axial vector fermionic currents $j^{\lambda}\equiv \bar{\psi}\gamma^{\lambda}\psi$ and $a^{\lambda}\equiv \bar{\psi}\gamma^{\lambda}\gamma^{5}\psi$, which are coupled to torsion via the matter Lagrangian density
\begin{equation}
\label{Lagnonmin}
\mathcal{L}_{\rm fermions}=\tilde{\mathcal{L}}_{\rm Dirac}+\alpha_{1}\bold{T}\cdot\bold{j}+\alpha_{2}\breve{\bold{T}}\cdot\bold{a} \ ,
\end{equation}
where $T^{\lambda}\equiv T^{\nu\lambda}_{\;\;\;\nu}$ is the trace vector part of torsion and $\alpha_1,\alpha_2$ some constants. For $\alpha_{1}=0$ and $\alpha_{2}=3\hbar/2$  we recover the case of the minimal coupling to torsion analyzed in the previous section. The extended Dirac equation in this case reads
\begin{equation}
\label{neweq:DiracFockIvaneckoHeisenberg0}
i\hbar \gamma^{\mu}\tilde{D}_{\mu}\psi-m\psi =-\alpha_{1}T^{\lambda}\gamma_{\lambda}\psi-
\alpha_{2} \breve{T}^{\lambda}\gamma_{\lambda}\gamma^{5}\psi \,.
\end{equation}

This model for the free fermion in a RC space-time geometry has parity symmetry, and can be extended into a family of parity-breaking models. To include parity-breaking terms (for a more detailed account of parity violation in the general framework of Poincare theories of gravity see the recent work \cite{Obukhov:2020zal}) in the Lagrangian density (which are expected to be relevant in the early Universe due to the matter-antimatter asymmetry), we consider the additional couplings  $\bold{T}\cdot\bold{a}$ and $\bold{\breve{T}}\cdot\bold{j}$, which yields the new Lagrangian density
\begin{equation}
\label{general Lagferm}
\mathcal{L}_{\rm fermions}=\tilde{\mathcal{L}}_{\rm Dirac}+(\alpha_{1}\bold{T}+\beta_{2}\breve{\bold{T}})\cdot\bold{j}+(\alpha_{2}\breve{\bold{T}}+\beta_{1}\bold{T})\cdot\bold{a} \ ,
\end{equation}
with new coefficients $\beta_1,\beta_2$. The corresponding generalized Dirac equation is given by
\begin{eqnarray}
\label{neweq:DiracFockIvaneckoHeisenberg}
i\hbar \gamma^{\mu}\tilde{D}_{\mu}\psi-m\psi &=&-\left( \alpha_{1}T^{\lambda}+\beta_{2}\breve{T}^{\lambda}\right) \gamma_{\lambda}\psi \nonumber \\
&&-\left( \alpha_{2} \breve{T}^{\lambda}+\beta_{1}T^{\lambda}\right) \gamma_{\lambda}\gamma^{5}\psi\,,
\end{eqnarray}
for spinors and
\begin{eqnarray}
\label{neweq:DiracFockIvaneckoHeisenbergadjoint}
i\hbar (\tilde{D}_{\mu}\bar{\psi})\gamma^{\mu}+m\bar{\psi} &=&-\left( \alpha_{1}T^{\lambda}+\beta_{2}\breve{T}^{\lambda}\right) \bar{\psi}\gamma_{\lambda} \nonumber \\
&&-\left( \alpha_{2} \breve{T}^{\lambda}+\beta_{1}T^{\lambda}\right)\bar{\psi}\gamma_{\lambda}\gamma^{5} \ ,
\end{eqnarray}
for the adjoint spinors.

To estimate the new physics involved in this model we will take again the zero-curvature  limit in order to identify the effects of torsion and have a qualitative notion of its consequences in the context of beyond the standard model of particle physics interactions. In this limit, the Dirac equation (\ref{neweq:DiracFockIvaneckoHeisenberg}) becomes
\begin{eqnarray}
\label{neweq:DiracFockIvaneckoHeisenbergflat}
i\hbar \gamma^{\mu}\partial_{\mu}\psi-m\psi &=&-\left( \alpha_{1}T^{\lambda}+\beta_{2}\breve{T}^{\lambda}\right) \gamma_{\lambda}\psi \nonumber \\
&&-\left( \alpha_{2} \breve{T}^{\lambda}+\beta_{1}T^{\lambda}\right) \gamma_{\lambda}\gamma^{5}\psi \ ,
\end{eqnarray}
As in the minimally coupled case, one can in principle consider different ansatze for the background torsion depending on whether one is interested, for instance, in gravitational wave astronomy (setting a dynamic, harmonic torsion), or in simple models of the RC geometry around spherical compact objects (setting a spherically symmetric ansatz). For free fermionic spinors we try again the solutions $\psi=\psi(\vec{r})e^{-iEt/\hbar}=e^{-i(Et-\vec{p}\cdot \vec{r})/\hbar}$. Accordingly, we have the following time-independent Dirac equation
\begin{eqnarray}
\label{neweq:DiracFockIvaneckoHeisenbergflattimeind}
&&-i\hbar \gamma^{k}\partial_{k}\psi+\Big[m-\left( \alpha_{1}T^{\lambda}+\beta_{2}\breve{T}^{\lambda}\right) \gamma_{\lambda} \nonumber \\
&& \qquad -
\Big(\alpha_{2} \breve{T}^{\lambda}+\beta_{1}T^{\lambda}\Big) \gamma_{\lambda}\gamma^{5}\Big]\psi(\vec{r})=\gamma^{0}E\psi(\vec{r}) \ .
\end{eqnarray}
Using the properties of the matrices $\gamma_{k}=-\gamma^{k}$, $\gamma_{k}\gamma^{5}$ and $\gamma^{0}$, we can also write this equation via the Hamiltonian matrix
\begin{equation}
\hat{H}\psi(\vec{r})=E\psi(\vec{r}) \ ,
\end{equation}
which explicitly reads as
\begin{eqnarray}
&&
\begin{pmatrix}
m-(t^{0}-\vec{\tau}\cdot\vec{\sigma}) & \;\;\vec{\sigma}\cdot\hat{\vec{p}}-(\tau^{0}-\vec{t}\cdot\vec{\sigma}) \\
-\vec{\sigma}\cdot\hat{\vec{p}}+(\tau^{0}-\vec{t}\cdot\vec{\sigma}) & \;\; m+(t^{0}-\vec{\tau}\cdot\vec{\sigma})
\end{pmatrix}\begin{pmatrix}
\psi_{I}\\
\psi_{II}
\end{pmatrix} \nonumber \\
&& \qquad \qquad \qquad =\begin{pmatrix}
E&0 \\
0&-E
\end{pmatrix}\begin{pmatrix}
\psi_{I} \\
\psi_{II}
\end{pmatrix}  \ ,
\end{eqnarray}
where $\hat{\vec{p}}=-i\hbar\vec{\nabla}$ is the 3-momentum operator, and we have introduced the following notation for the torsion quantities
\begin{equation} \label{eq:torquan}
t^{\nu}\equiv \alpha_{1}T^{\nu}+\beta_{2}\breve{T}^{\nu},\qquad \tau^{\lambda}\equiv \alpha_{2} \breve{T}^{\lambda}+\beta_{1}T^{\lambda} \ .
\end{equation}

Alternatively, this system can also be written in the more convenient way
\begin{eqnarray}
&&
\begin{pmatrix}
m-E & \;\;\vec{\sigma}\cdot\hat{\vec{p}} \\
-\vec{\sigma}\cdot\hat{\vec{p}} & \;\; m+E
\end{pmatrix}\begin{pmatrix}
\psi_{I}\\
\psi_{II}
\end{pmatrix}
	\nonumber \\
&&  \qquad =\begin{pmatrix}
t^{0}-\vec{\tau}\cdot\vec{\sigma} & \;\;\tau^{0}-\vec{t}\cdot\vec{\sigma} \\
-\tau^{0}+\vec{t}\cdot\vec{\sigma} & \;\; -t^{0}+\vec{\tau}\cdot\vec{\sigma}
\end{pmatrix}\begin{pmatrix}
\psi_{I} \\
\psi_{II}
\end{pmatrix} \ ,
\end{eqnarray}
which highlights the fact that the matrix on the right-hand side contains the geometrical effects due to torsion, including spin-torsion interactions of both parity-breaking and parity-preserving types.

The eigenvalue problem above is a system of two coupled equations for the 2-spinors $\psi_{I}$ and $\psi_{II}$. To solve it we use the general form of the spinor $\psi(\vec{r})=\chi e^{i\vec{p}\cdot\vec{r}/\hbar}$ and the properties of Pauli matrices, so that the first of these equations can be written as
\begin{eqnarray}
&& \begin{pmatrix}p_{3}+t^{3}-\tau^{0} & \;\;p_{1}+t^{1}-i(p_{2}+t^{2})
 \\
p_{1}+t^{1}+i(p_{2}+t^{2}) & \;\;-p_{3}-t^{3}-\tau^{0}
\end{pmatrix}\begin{pmatrix}
\chi^{II}_{1}\\
\chi^{II}_{2}
\end{pmatrix}
	\nonumber \\
&& \quad =\begin{pmatrix}
E-m+t^{0}-\tau^{3}&-(\tau^{1}-i\tau^{2}) \\
-(\tau^{1}+i\tau^{2})&E-m+t^{0}+\tau^{3}
\end{pmatrix}\begin{pmatrix}
\chi^{I}_{1}\\
\chi^{I}_{2}
\end{pmatrix} .
\end{eqnarray}
Now let us consider the two orthogonal spin up/down solutions for the particle:  $\chi^{I}=\begin{pmatrix}
1\\
0
\end{pmatrix}$ and $\chi^{I}=\begin{pmatrix}
0\\
1
\end{pmatrix}$, and obtain the corresponding 4-spinor solutions. In the first (spin up) case, we get the system of equations
\begin{eqnarray}
&&(p_{3}+t^{3}-\tau^{0})\chi^{II}_{1}+(p_{1}+t^{1}-i(p_{2}+t^{2}))\chi^{II}_{2} \nonumber \\
&& \qquad \qquad \qquad =E-m+t^{0}-\tau^{3} ,
\end{eqnarray}
\begin{eqnarray}
&&(p_{1}+t^{1}+i(p_{2}+t^{2}))\chi^{II}_{1}+(-p_{3}-t^{3}-\tau^{0})\chi^{II}_{2} \nonumber \\
&& \qquad \qquad \qquad  =-(\tau^{1}+i\tau^{2}) ,
\end{eqnarray}
and therefore we find the solution $\chi=\begin{pmatrix}
1\\
0\\
\chi^{II}_{1}\\
\chi^{II}_{2}
\end{pmatrix}$, with
\begin{widetext}
\begin{equation}
\chi^{II}_{1}=\dfrac{(E-m+t^{0}-\tau^{3})(-p_{3}-t^{3}-\tau^{0})+(p_{1}+t^{1}-i(p_{2}+t^{2}))(\tau^{1}+i\tau^{2})}{(p_{3}+t^{3}-\tau^{0})(-p_{3}-t^{3}-\tau^{0})-(p_{1}+t^{1}-i(p_{2}+t^{2}))(p_{1}+t^{1}+i(p_{2}+t^{2}))},
\end{equation}
\begin{equation}
\chi^{II}_{2}=\dfrac{-(\tau^{1}+i\tau^{2})(p_{3}+t^{3}-\tau^{0})-(p_{1}+t^{1}+i(p_{2}+t^{2}))(E-m+t^{0}-\tau^{3})}{(p_{3}+t^{3}-\tau^{0})(-p_{3}-t^{3}-\tau^{0})-(p_{1}+t^{1}-i(p_{2}+t^{2}))(p_{1}+t^{1}+i(p_{2}+t^{2}))} \ .
\end{equation}
\end{widetext}

Note that, in the vanishing-torsion (Minkowski) limit we obtain $\chi_1^{II}=
\tfrac{p_{3}}{E+m}$, $\chi_2^{II}=\tfrac{p_{1}+ip_{2}}{E+m}$,  which is exactly the 4-spinor solution corresponding to the free fermion, spin up state, with $E^{2}=p^{2}+m^{2}$. As for the second (spin down) case,  we obtain the system
\begin{eqnarray}
&&(p_{3}+t^{3}-\tau^{0})\chi^{II}_{1}+(p_{1}+t^{1}-i(p_{2}+t^{2}))\chi^{II}_{2} \nonumber \\
&& \qquad \qquad \qquad=-(\tau^{1}-i\tau^{2}) ,
\end{eqnarray}
\begin{eqnarray}
&&(p_{1}+t^{1}+i(p_{2}+t^{2}))\chi^{II}_{1}+(-p_{3}-t^{3}-\tau^{0})\chi^{II}_{2} \nonumber \\
&& \qquad \qquad \qquad=E-m+t^{0}+\tau^{3} \ ,
\end{eqnarray}
and therefore we find the solution $\chi=\begin{pmatrix}
0\\
1\\
\chi^{II}_{1}\\
\chi^{II}_{2}
\end{pmatrix}$, with
\begin{widetext}
\begin{equation}
\chi^{II}_{1}=\dfrac{-(\tau^{1}-i\tau^{2})(-p_{3}-t^{3}-\tau^{0})-(p_{1}+t^{1}-i(p_{2}+t^{2}))(E-m+t^{0}+\tau^{3})}{(p_{3}+t^{3}-\tau^{0})(-p_{3}-t^{3}-\tau^{0})-(p_{1}+t^{1}-i(p_{2}+t^{2}))(p_{1}+t^{1}+i(p_{2}+t^{2}))}
\end{equation}
\begin{equation}
\chi^{II}_{2}=\dfrac{(E-m+t^{0}+\tau^{3})(p_{3}+t^{3}-\tau^{0})+(p_{1}+t^{1}+i(p_{2}+t^{2}))(\tau^{1}-i\tau^{2})}{(p_{3}+t^{3}-\tau^{0})(-p_{3}-t^{3}-\tau^{0})-(p_{1}+t^{1}-i(p_{2}+t^{2}))(p_{1}+t^{1}+i(p_{2}+t^{2}))} \ .
\end{equation}
\end{widetext}
Again we have the correct Minkowski limit, $ \chi_1^{II}=
\tfrac{p_{1}-ip_{2}}{E+m}$, and $\chi_2^{II}=\tfrac{-p_{3}}{E+m}$, describing the free particle, spin down state. Note that, proceeding in a similar manner, we could derive the corresponding expressions for the 4-spinor solutions associated to the anti-fermion in the spin up/down states.

To simplify further our analysis let us consider the ansatz for the torsion components $t^{\mu}=(0,t^{1},t^{2},t^{3})$, and $\tau^{\mu}=(0,0,0,\tau)$ in Eq. (\ref{eq:torquan}). The spin up particle solution is then given by
\begin{equation}
\psi=\begin{pmatrix}
1\\
0\\
\dfrac{p^{\rm eff}_{3}}{E+m_{\rm eff}}\\
\dfrac{p_{1}^{\rm eff}+ip_{2}^{\rm eff}}{E+m_{\rm eff}}
\end{pmatrix}e^{i(\vec{p}\cdot\vec{r}-Et)/\hbar} \ ,
\end{equation}
where
\begin{equation} \label{eq:Esquared}
E^{2}=p^{2}_{\rm eff}+m^{2}_{\rm eff} \ ,
\end{equation}
with the definitions
\begin{eqnarray}
p_{k}^{\rm eff} &\equiv& p_{k}+t^{k}  \label{eq:pks1} \\
p^{2}_{\rm eff}&\equiv&(p_{1}+t^{1})^{2}+(p_{2}+t^{2})^{2}+(p_{3}+t^{3})^{2} \label{eq:pks2} \\
m_{\rm eff}&\equiv& m+\tau \label{eq:pks3} \ .
\end{eqnarray}
Analogously, for the spin down particle we get the solution
\begin{equation}
\psi=\begin{pmatrix}
0\\
1\\
\dfrac{p^{\rm eff}_{1}+ip^{\rm eff}_{2}}{E+m_{\rm eff}}\\
\dfrac{-p^{\rm eff}_{3}}{E+m_{\rm eff}}
\end{pmatrix}e^{i(\vec{p}\cdot\vec{r}-Et)/\hbar} \ ,
\end{equation}
where Eqs. (\ref{eq:Esquared})--(\ref{eq:pks2}) still hold but  now the effective mass in Eq.  (\ref{eq:pks3}) becomes $m_{\rm eff}\equiv m-\tau$. Therefore, two different energy levels are obtained for the spin up and spin down states. The energy of the anti-aligned state with respect to the direction of $\vec{\tau}$ is lower than the aligned state. These two possible energy states, $E_{2}^2=p_{\rm eff}^2+(m+\tau)^2$ and $E_{1}^2=p_{\rm eff}^2+(m-\tau)^2$, correspond to the energy transition
\begin{equation}
h\nu= E_{2}-E_{1}=\dfrac{4m\tau}{E_1+E_2} \ ,
\end{equation}
which in the reference frame of the particle reads
\begin{equation}
h\nu= \dfrac{4m\tau}{\left[\vec{t}^{2}+(m+\tau)^{2}\right]^{1/2}+\left[\vec{t}^{2}+(m-\tau)^{2}\right]^{1/2}} \ ,
\end{equation}
where $\vec{t}^{2}\equiv (t^{1})^{2}+(t^{2})^{2}+(t^{3})^{2}$ can be written simply as $t^2$ assuming that $\vec{t}$ is aligned in any of the spatial axis directions of the reference system of coordinates. We recall that the torsion functions $t^{\mu}$ and $\tau^{\mu}$ are constructed from the torsion trace vector and axial vectors and depend on the (parity-preserving) $(\alpha_{1},\alpha_{2})$  and (parity-breaking) $(\beta_{1},\beta_{2})$ coupling parameters.

The bottom line of this section is that parity-breaking effects can arise in a RC space-time from the interaction of fermions with a background torsion field via non-minimal couplings. These effects include the prediction of well-defined frequencies that a free fermion can absorb or emit in order to make transitions between the predicted two energy levels that arise depending on the spin orientation with respect to external torsion quantities. The signature of parity breaking might also be present in the radiated field itself.

\section{Non-minimal couplings in the Einstein-Cartan theory} \label{sec:V}

This framework could be further specified by considering the EC theory.
The corresponding Lagrangian then reads
\begin{equation} \label{eq:actionEC}
S_{\rm EC}=\dfrac{1}{2\kappa^2} \int d^4x \sqrt{-g}R(\Gamma) + \int d^4x \sqrt{-g}\,\mathcal{L}_{\rm fermions}\,,
\end{equation}
where the Ricci scalar of the independent connection, $R(\Gamma)$, with $\Gamma \equiv \Gamma_{\mu \nu}^{\lambda}$, can be related to the one constructed with the metric-compatible connection, $\tilde{R}(\tilde{\Gamma})$, via an expression of the form
\begin{equation}
R\sim\tilde{R}-4\tilde{\nabla}_{\alpha}T^{\alpha}-\dfrac{1}{3}T^{\lambda}T_{\lambda}+\dfrac{1}{24}\breve{T}^{\lambda}\breve{T}_{\lambda}+\dfrac{1}{2}\bar{T}_{\mu\nu\rho}\bar{T}^{\mu\nu\rho}.
\end{equation}
Inserting this in the action (\ref{eq:actionEC}) and taking the fermionic Lagrangian defined in Eq. (\ref{Lagnonmin}), the corresponding Cartan equations become
\begin{equation}
T^{\mu}\sim\kappa^{2}\alpha_{1}j^{\mu} \,, \qquad \breve{T}^{\mu}\sim\kappa^{2}\alpha_{2}a^{\mu} \ .
\end{equation}
Re-inserting these expressions in the (\ref{Lagnonmin}) we obtain effective vector-vector contact interactions besides the usual well-known axial-axial (spin-sin) interaction (Hehl-Data term) as
\begin{equation}
\mathcal{L}_{\rm fermions}\sim\tilde{\mathcal{L}}_{\rm Dirac}+\dfrac{\kappa^{2}}{3}(\alpha_{1})^{2} {\bf j}\cdot {\bf j}-\dfrac{\kappa^{2}}{24}(\alpha_{2})^{2} {\bf a}\cdot {\bf a} \ .
\end{equation}

The corresponding Dirac equation can be written as
\begin{eqnarray}
\label{new2eq:DiracFockIvaneckoHeisenberg}
i\hbar \gamma^{\mu}\tilde{D}_{\mu}\psi-m\psi &=&
\dfrac{\kappa^{2}\alpha_{2}^{2}}{12}(\bar{\psi}\gamma^{\lambda}\gamma^{5}\psi)\gamma_{\lambda}\gamma^{5}\psi \nonumber \\
&&-\kappa^{2}\dfrac{2\alpha_{1}}{3}(\bar{\psi}\gamma^{\lambda}\psi)\gamma_{\lambda}\psi \,.
\end{eqnarray}
As in the usual Dirac-Hehl-Data equation, under charge conjugation operation $\psi\rightarrow\psi^{ch}$ one obtains different dynamics for the $\psi^{ch}$ representing anti-fermions. If we use instead the Lagrangian in (\ref{general Lagferm}), then the Cartan equations are
\begin{equation}
T^{\mu}\sim\kappa^{2}(\zeta_{1}j^{\mu}+\zeta_{2}a^{\mu})\,, \qquad \breve{T}^{\mu}\sim\kappa^{2}(\theta_{1}j^{\mu}+\theta_{2}a^{\mu}) \ ,
\end{equation}
where $\zeta_{i},\theta_{i}$ ($i=1,2$) are constants, and the resulting Dirac equation, after substitution in (\ref{neweq:DiracFockIvaneckoHeisenberg}), includes parity-breaking and C-breaking cubic terms.

The EC theory can also be extended to include the parity-breaking Holst term \cite{Holst:1995pc}, encapsulating additional vector-axial (contact) self-interactions. The extended action is thus
\begin{eqnarray} \label{eq:actionECHolst}
S_{\rm EC}&=&\dfrac{1}{2\kappa^2} \int d^4x \sqrt{-g}R(\Gamma)
+ \int d^4x \sqrt{-g}\,\mathcal{L}_{\rm fermions}
	\nonumber \\
&&	+\dfrac{1}{2\gamma\kappa^2} \int d^4x \sqrt{-g}\epsilon^{\alpha\beta\mu\nu}R_{\alpha\beta\mu\nu}  \ ,
\end{eqnarray}
where $\gamma$ is the Barbero-Immirzi parameter and the parity-breaking Holst term $\epsilon^{\alpha\beta\mu\nu}R_{\alpha\beta\mu\nu}$ can be expressed as
\begin{equation}
\epsilon^{\alpha\beta\mu\nu}R_{\alpha\beta\mu\nu}\sim-\tilde{\nabla}_{\alpha}\breve{T}^{\alpha}-\dfrac{1}{3}\breve{T}^{\lambda}T_{\lambda}+\dfrac{1}{2}\epsilon^{\alpha\beta\mu\nu}\bar{T}^{\lambda}_{\;\;\alpha\beta}\bar{T}_{\lambda\mu\nu} \ .
\end{equation}
The generalized Cartan equations  become
\begin{eqnarray}
T^{\mu} & \sim & \kappa^{2}\dfrac{3\gamma}{1+\gamma^{2}}(\alpha_{1}\gamma j^{\mu}+a^{\mu}) \,, \\
\breve{T}^{\mu} &\sim &\kappa^{2}\dfrac{3\gamma}{1+\gamma^{2}}(\alpha_{1}j^{\mu}-\alpha_{2}\gamma a^{\mu}) \,.
\end{eqnarray}
Then, by choosing the fermionic Lagrangian in Eq. (\ref{Lagnonmin}), one obtains a generalized Dirac equation and Lagrangian with vector-vector, axial-axial and parity-breaking vector-axial (contact) self-interactions. As in the model-independent approach of the previous section, all these cases of non-minimal couplings in specific gravitational models (EC and its Holst extension) yield interesting and quite relevant physics (C and P symmetry-breaking, beyond standard model interactions, etc) that in principle can be observationally probed, upon the computation of the energy levels of fermionic systems and searching for its signatures using advanced spectrographs.

\section{Discussion and conclusion}  \label{sec:VI}

Experimental constraints on the minimal and non-minimal coupling of spinors with torsion using high-precision data has been a topic of interest in the literature since quite a long time ago \cite{Kostelecky,Heckel:2008hw,Lehnert:2013jsa}. In this sense, the results derived in the present paper may have a physical impact at  several levels. First, by generalizing these results to the case of bound states of electrons within atoms and molecules, and also to bound states of nucleons within atomic nuclei, one opens up the possibility of detecting the effects of torsion in the strong gravitational regime via the measurement of  spectral lines and searching for new fine structures, for instance, using ultra-sensitive spectrographs.  Therefore, there could be astrophysical spectral signatures of torsion waiting to be discovered around intense gravitational fields of neutron stars, or even in X-ray binaries, where one of the objects is a black hole candidate surrounded by an accretion disk. For a given bound system, a specific initial energy level could be chosen such that the transitions from this level into the two (Zeeman-like) lower levels due to the torsion-spin interaction could be searched for. Indeed, this could be carried out for different values of the predicted torsion in the emission regions according to different gravitational theories accommodating torsion effects.

There are further examples of astrophysical interest where torsion effects into the physics of fermions/anti-fermions could be observationally detected. A particular case is that of polarized fermion/anti-fermion pairs produced within quantum field theory, for instance in the space-time torsion around black holes. After their subsequent annihilation, the generated photons would have different energies depending on whether the spins of the fermions in the pair are up or down with respect to the background torsion. In the particle pair production in general, there are three possible scenarios: i) particle and antiparticle are anti-aligned with each other, ii) both are with their spins up (i.e. aligned to torsion), iii) both are with their spins down. These correspond to the three energy/mass levels that the (initial) photons can generate via pair production, with the first case having an  intermediate energy, the second the highest, and the third the lowest. In the very early Universe, such effects would depend on the temperature of the quark-gluon-lepton-photon plasma, while in Hawking's radiation the outgoing and ingoing energy flux through the event horizon would be spin state-dependent. In particular, the (outflux) energy loss via particles (or antiparticles) aligned with the background torsion would be more efficient.

These effects could also be sought for in the emission of gamma rays in high energy astrophysical environments driven by strong gravitational fields. An observable signature of the background torsion could be then obtained by comparing the measured flux spectra with the detailed theoretical prediction of the emission curves. In the most general case the theory suggests that the radiated flux should result from the superposition of the three possibilities above peaked at the characteristic nearby frequencies, corresponding to the three possible values of the energy of the (annihilating) pair. Disregarding complex environmental effects and significant changes in the background torsion (in the typical scales of the emission region), the emission curves would resemble emission lines very close together (a kind of hyperfine structure). In the more conventional case of pair production with the fermion and anti-fermion anti-aligned with respect to each other, the theory predicts a specific characteristic frequency, determined by the particle's mass, the torsion field, and the temperature of the emission region, slightly deviated from the corresponding predicted frequency when torsion is absent.

Beyond astrophysics and cosmology, torsion-induced effects of the kind studied in this work might be found in the field of condensed matter physics. Such is the case of the interaction of a Dirac fermion  with the torsion of a sea of vacuum fermion-condensates, provided that the latter has a non-zero expectation value. This way, fermions in vacuum would have a different effective mass according to the relative orientation of the fermionic spin with respect to the background vacuum axial torsion, which could be tested in laboratories, putting bounds on the predicted effects. While these tests do not require strong gravitational fields the challenge lies on reproducing the conditions of fermionic vacuum condensates in the laboratory.

Another possibility would be the existence of a continuous and smooth phase transition for a Bose-Einstein system as a superconducting fluid/material in a space-time background with torsion. Cooper pairs of anti-aligned fermions in bound states are required in the BCS model of superconductivity and in general Bose-Einstein condensates, and since the effective mass of the pair would depend on the interaction with the torsion background, the effective spin-zero bosonic field due to the ensemble of Cooper pairs would have a differential effective mass powered by the relative strength of torsion. If such were the case, then this would have a non-negligible impact on the superconducting and superfluid phases in the interiors of neutron stars and hypothetical quark and strange stars, with consequences on the macroscopic predictions of stellar models (mass-radius relations, moment of inertia, etc). Moreover, the stability of Cooper pairs might be strongly perturbed as the torsion increases above a certain threshold, since the background torsion axial vector along a well defined direction can act exactly as an external magnetic field does in paramagnetic materials, i.e., above a certain critical value of the external field a significant number of large clusters of ``aligned'' spins are developed (and percolating the whole system) and the material is magnetized. The spin-spin interaction that naturally exists in a system with spins is analogous to a thermal-like interaction (increasing temperature tends to rise the entropy, and generate a random distribution of spins), while the external field tends to counteract the random distribution of spins, by establishing gradually a more ordered state. Therefore, torsion can also act as an external field driving a phase transition in a macroscopic system of microphysical components with spin, magnetizing the material, with the emergence of a macroscopic (intrinsic) spin. In that sense, the superfluid/superconducting phase of the BCS models could suffer a phase transition for sufficiently strong ``external'' space-time torsion, inside ultra-dense compact objects. These topics deserve a much more careful analysis, since they evolve very complicated physics of the interiors of neutron stars and related objects.

Further avenues of research are those involving laboratory tests of the space-time torsion near Earth. In perfect analogy with the magnetic spin resonance, one could design a torsion-spin resonance. In this effect, an external torsion field generates the splitting of energy levels (Zeeman-like effect) in an appropriate material sample, while a time-varying current produces an electromagnetic wave which suffers a measurable absorption once the resonance frequency is achieved matching the energy gap. Therefore, the indirect detection of torsion would be achieved by the measurement of absorption (decrease in intensity) of the electromagnetic wave interacting with the material sample, once the resonance frequency is achieved. For free fermions we saw that the predicted frequency (in the particle's frame) does not depend on the fermionic mass, only on the background torsion. If  torsion has a magnitude of about $10^{-16}-10^{-15}$ m$^{-1}$ then we get an estimated resonance frequency around $1-10$ nHz, which corresponds to resonance in the radio band.

Finally, regarding the non-minimal couplings within specific gravity models such as the EC one and its extension with the Holst term,  one also obtains generalized Dirac equations and Lagrangians with vector-vector, axial-axial and parity-breaking vector-axial (contact) self-interactions. These might be relevant inside compact objects like neutron, quark, and strange (quark) stars and also in the early Universe. If the coupling constants are taken to be dynamical scalar fields, then this scenario leads naturally to the idea of parity-breaking phase transitions for matter under extreme conditions, induced by the torsion-fermion currents couplings. We also see that EC gravity plus Holst with $T\cdot j$ and $\breve{T}\cdot a$ couplings can be made equivalent to the usual EC theory with $T\cdot j$ and $\breve{T}\cdot a$ plus (parity-breaking) $T\cdot a$ and $\breve{T}\cdot j$ couplings.

To conclude, the results obtained in this paper open up new avenues for testing non-Riemannian geometries with torsion using splitting of energy levels in both minimally and non-minimally coupled fermions to the background torsion in a variety of astrophysical/cosmological environments.  Further work along these lines is currently underway.

\section*{Acknowledgments}
We thank A. Delhom for useful comments and insights. FC is funded by the Funda\c{c}\~ao para a Ci\^encia e a Tecnologia (FCT, Portugal) doctoral grant No.PD/BD/128017/2016.
FSNL acknowledges support from the FCT Scientific Employment Stimulus contract with reference
CEECIND/04057/2017.
DRG is funded by the \emph{Atracci\'on de Talento Investigador} programme of the Comunidad de Madrid (Spain) No. 2018-T1/TIC-10431, and acknowledges further support from the Ministerio de Ciencia, Innovaci\'on y Universidades (Spain) project No. PID2019-108485GB-I00/AEI/10.13039/501100011033, the Spanish project No. FIS2017-84440-C2-1-P (MINECO/FEDER, EU), the project PROMETEO/2020/079 (Generalitat Valenciana), and the Edital 006/2018 PRONEX (FAPESQ-PB/CNPQ, Brazil) Grant No. 0015/2019. The authors also acknowledge funding from FCT Projects No. UID/FIS/04434/2020, No. CERN/FIS-PAR/0037/2019 and No. PTDC/FIS- OUT/29048/2017.
FC thanks the hospitality of the Department of Theoretical Physics and IPARCOS of the Complutense University of Madrid, where part of this work was carried out.  This article is based upon work from COST Action CA18108, supported by COST (European Cooperation in Science and Technology).



\end{document}